\documentclass[aps,prd,longbibliography,twocolumn,amsmath,amssymb,noeprint,floatfix]{revtex4-2}

\usepackage{graphicx}
\usepackage{dcolumn}
\usepackage{bm}
\usepackage{latexsym}
\usepackage{slashed}
\usepackage[colorlinks=true,linkcolor=blue,citecolor=blue]{hyperref}
\usepackage{color} 

\newcommand{\Rco}{$\mathcal R_{\slashed E}$}

\begin{document}

\title{First-principles Fermi acceleration in magnetized turbulence}

\author{Martin Lemoine} 
\affiliation{Institut d'Astrophysique de Paris,\\
CNRS -- Sorbonne Universit\'e, \\
98 bis boulevard Arago, F-75014 Paris, France}

\date{\today}

\begin{abstract}  
This work provides a concrete implementation of E. Fermi's model of particle acceleration in magnetohydrodynamic (MHD) turbulence, connecting the rate of energization to the gradients of the velocity of magnetic field lines, which it characterizes within a multifractal picture of turbulence intermittency. It then derives a transport equation in momentum space for the distribution function. This description is shown to be substantiated by a large-scale numerical simulation of strong MHD turbulence. The present, general framework can be used to model particle acceleration in a variety of environments.
\end{abstract}

\maketitle

\noindent\emph{Introduction--\,}
Particle energization through scatterings off inhomogeneous, random moving structures is a universal process~\cite{1949PhRv...75.1169F,1954ApJ...119....1F}, which has stirred considerable interest in various branches of physics: primarily astrophysics, with applications ranging from solar flares~\cite{2012SSRv..173..535P} to more remote phenomena involving plasmas in extreme conditions, {\it e.g.}~\cite{2011ApJ...739...66T}, but also statistical plasma physics~\cite{1972SvPhU..14..549Z} or high energy density physics~\cite{2018JPlPh..84f9008B}. Remarkably, the two papers of E. Fermi~\cite{1949PhRv...75.1169F,1954ApJ...119....1F} represent the first concrete scenarios for the origin of non-thermal particles in the Universe. While the literature has placed significant emphasis on acceleration at shock fronts, numerical experiments have demonstrated that stochastic acceleration can be efficient~\cite{2003ApJ...597L..81D,2020ApJ...894..136T,2022ApJ...928...25P}, notably so at large turbulent Alfv\'en velocity~\cite{17Zhdankin,18Comisso,2019ApJ...886..122C,2020ApJ...893L...7W,2021ApJ...922..172Z}, in the sense that it produces extended, hard powerlaw distributions of suprathermal particles. Besides, the stochastic Fermi process assuredly plays a role in the vicinity of shock fronts~\cite{2017PhRvL.119j5101M,2021PNAS..11826764T,2022ApJ...926..109N}, just as it seemingly controls part of the energization in reconnection environments~\cite{2006Natur.443..553D,2019ApJ...879L..23G}.

While the overall scenario is commonly pictured as originally formulated by E. Fermi -- a sequence of discrete, point-like interactions between a particle and infinitely massive, perfectly conducting plasma clouds -- its implementation in a realistic turbulent context has remained a challenge~\cite{1993PhyU...36.1020B,PhysRevLett.92.040601,PhysRevLett.111.195001}, to the extent that phenomenological applications rely on a Fokker-Planck model parameterized by a momentum diffusion coefficient. 

The present Letter proposes a novel approach to this problem and formulates a  transport equation to describe the evolution of the distribution function in momentum space. It is first shown that particle momenta obey a continuous-time random walk (CTRW), whose random force scales as the gradients of the velocity of magnetic field lines, coarse-grained on a scale comparable to the particle gyroradius $r_{\rm g}\equiv pc/eB$ ($p$ momentum, $B=\vert\boldsymbol{B}\vert$ with $\boldsymbol{B}$ magnetic field). A key observation is that those gradients are subject to intermittency on small length scales. Hence, the random forces are neither Gaussian, nor white noise in time and consequently, the random walk deviates from Brownian motion, just as the transport equation, which is derived here from known properties of CTRW, differs from Fokker-Planck. This equation is characterized by the statistics of velocity gradients, which are captured via a multifractal description of turbulence intermittency. This framework is eventually shown to reproduce the time- and momentum-dependent Green functions obtained by tracking a large number of test particles in a large-scale MHD simulation. The present formalism thus provides a successful implementation of stochastic Fermi acceleration in realistic, collisionless MHD turbulence.
 \smallskip

\noindent\emph{A (continuous-time) random walk picture--\,} To evaluate energy gains/losses in the original Fermi model, it proves convenient to boost to the scattering center frame where the motional electric field $\boldsymbol{E}$ vanishes. The generalization of that model to a continuous random flow similarly tracks the particle momentum in the instantaneous (here, non-inertial) frame \Rco\ in which $\boldsymbol{E}$ vanishes~\cite{2019PhRvD..99h3006L,PhysRevD.104.063020}, which, in ideal MHD, moves at velocity $\boldsymbol{v_E}=c\,\boldsymbol{E}\times\boldsymbol{B}/B^2$. In that frame, momentum gains or losses scale in direct proportion to the (lab frame) spatio-temporal gradients of the velocity field $\boldsymbol{v_E}$, as expressed by $\Gamma_{\rm acc}$, $\Gamma_{\parallel}$ and $\Gamma_{\perp}$ below. In detail, the momentum $p$ of particles with gyroradius $r_{\rm g}\ll \ell_{\rm c}$ ($\ell_{\rm c}$  coherence length of the turbulence) evolves as 
\begin{equation}
\dot{p}\,=\,p\left\{\Gamma_{\rm acc} + \Gamma_{\parallel} + \Gamma_\perp\right\}\,,
\label{eq:deltap}
\end{equation}
with $\Gamma_{\rm acc} = -v^{-1}\,\mu\,\boldsymbol{b}\cdot\partial_t \boldsymbol{v_E}$ ($v$ particle velocity; $\mu = \boldsymbol{p}\cdot\boldsymbol{b}/p$  pitch-angle cosine with respect to the magnetic field direction $\boldsymbol{b}=\boldsymbol{B}/B$); $\Gamma_{\parallel} = -\,{\mu}^2\,\boldsymbol{b}\cdot\left(\boldsymbol{b}\cdot\boldsymbol{\nabla}\right)\boldsymbol{v_E}$ and $\Gamma_{\perp} = - \,\left(1-{\mu}^2\right)\left[\boldsymbol{\nabla}\cdot\boldsymbol{v_E}-\boldsymbol{b}\cdot\left(\boldsymbol{b}\cdot\boldsymbol{\nabla}\right)\boldsymbol{v_E}\right]/2$. 
For simplicity, the present work focuses on the sub-relativistic limit $v_E \ll c$. Equation~(\ref{eq:deltap}) -- more precisely, its relativistic limit -- has been shown to account for the bulk of acceleration in numerical simulations of collisionless turbulence~\cite{Bresci+22}, putting the present model on solid footing. It generalizes the two contributions originally identified by E. Fermi: interactions with moving magnetic mirrors are captured by $\Gamma_\perp$, while orbits in dynamic, curved magnetic field lines are represented by $\Gamma_\parallel$; the remaining term $\Gamma_{\rm acc}$ describes the effective gravity force associated to acceleration/deceleration of the field lines; of second order in $v_E/c$, it is subdominant in the sub-relativistic limit, unless $v\ll v_{\rm A}$.

All quantities in Eq.~(\ref{eq:deltap}) are understood to be coarse-grained on wavenumber scale $k_{\rm g}\sim r_{\rm g}^{-1}$ (length scales $l_{\rm g}\sim 2\pi\,r_{\rm g}$), where $r_{\rm g}=pc/eB$ denotes the particle gyroradius in the lab frame. This procedure filters out wavenumbers $k> k_{\rm g}$, whose contribution averages out over a gyro-orbit, to retain the larger scales that shape the velocity structures responsible for acceleration (in accord with the original Fermi picture).

Henceforth, Eq.~(\ref{eq:deltap}) is simplified to the symbolic $\dot p = p\,\Gamma_{l_{\rm g}}$, $\Gamma_{l_{\rm g}}$ representing an aggregate (random) force exerted by electromagnetic fields, coarse-grained on scale $l_{\rm g}$; order of unity factors related to $\mu$ are thus omitted; we also consider relativistic particles ($v\sim c$) to ease the discussion. For technical details concerning the modeling of this random process, see [Supp.~Mat.~A]. Separate now fluctuations from the mean, $\Gamma_l = \langle \Gamma_l\rangle + \delta \Gamma_l$,  the average carrying over the statistical realizations of the turbulent flow: $\langle\Gamma_l\rangle$ characterizes systematic heating, while the random $\delta\Gamma_l$ represents the diffusive part. If $\delta\Gamma_l$ were Gaussian distributed, and its time correlation function that of white noise, the process would describe Brownian motion, in one-to-one correspondence with a Fokker-Planck equation for the distribution function~\cite{1989fpem.book.....R}. As anticipated above, however, those random forces are neither Gaussian in amplitude, nor white noise in time: at small scales, they develop large powerlaw tails as a result of intermittency, while at large scales, the coherence time of the random force $\gtrsim l_{\rm g}/c$ cannot be regarded as infinitesimal. 

To obtain the transport equation, we first observe that the process $\dot p = p\,\Gamma_{l_{\rm g}}$ can be described as a CTRW: unlike discretized Brownian motion, which operates at a fixed and uniform time step, the random walk is here defined by the joint probability $\phi\left(p\vert p';\,t-t'\right)$ to jump from $p'$ to $p$ in time $\Delta t=t-t'$, with both $\Delta p=p-p'$ and $\Delta t$ regarded as random variables. Expectations are $\Delta t \sim l_{\rm g}(p')/c$ -- thus, a function of $p'$ -- and $\Delta \ln p \sim \Gamma_{l_{\rm g}}\Delta t$. We will assume $\Delta t$ to be exponentially distributed with mean waiting time $t_p\equiv l_{\rm g}(p')/c$ and $\Delta \ln\,p$ distributed as $\Gamma_{l_{\rm g}} l_{\rm g}/c$ [$l_{\rm g}=l_{\rm g}(p')$], see [Supp.~Mat.~A] for methodology. The random walk is then entirely defined by the statistics of the velocity gradients $\Gamma_{l_{\rm g}}$.  
\smallskip

\noindent\emph{Statistics of momentum jumps--\,} In turbulence theories, such  statistics are conveniently described within a multifractal analysis~\cite{Frisch..Parisi.85,1984JPhA...17.3521B,1991PhRvL..67.2299B}, which ascribes to each position $\boldsymbol{x}$ a local scaling exponent $h(\boldsymbol{x})$ for gradients on coarse-graining scale $l$, {\it viz.} 
\begin{equation}
\Gamma_l(\boldsymbol{x}) \,\sim\, \Gamma_{\ell_{\rm c}}(\boldsymbol{x})\, \left(l/\ell_{\rm c}\right)^{h(\boldsymbol{x})}\,,
\label{eq:deltaU}
\end{equation}
and which describes the set of locations $\boldsymbol{x}$ with index $h(\boldsymbol{x})$ as a fractal of dimension $d(h)$. The statistics of $\Gamma_l$ are thus entirely captured by the probability distribution function (p.d.f.) ${\textsf p}_{\Gamma_{\ell_{\rm c}}}$ of $\Gamma_{\ell_{\rm c}}$  and by the spectrum $d(h)$, since the probability of being at $\boldsymbol{x}$ in a set with exponent $h$ on scale $l$ evolves as the volume filling fraction $l^{D-d(h)}$ ($D$ number of spatial dimensions). The gradient $\Gamma_{\ell_{\rm c}}(\boldsymbol{x})$ on the coherence scale $\ell_{\rm c}$ is naturally modeled as a Gaussian variable with standard deviation $\sigma_{\rm c}\sim v_{\rm A}/\ell_{\rm c}$, where $v_{\rm A}$ denotes the Alfv\'en velocity of the turbulent component. The spectrum $d(h)$ can take different forms, the simplest being log-normal~\cite{2019JFM...867P...1D}, modern descriptions of the statistics of Els\"asser fields in MHD turbulence rather relying on log-Poisson models~\cite{2000PhPl....7.4889B,2003PhRvE..67f6302M,2004PhRvL..92s1102P,2015ApJ...807...39C}. We use the former log-normal form, as it provides a simple and satisfactory description of the statistics of the gradients of $\boldsymbol{v_E}$; see [Supp.~Mat.~B], which includes Refs.~\cite{1994PhRvL..72..336S,1994PhRvL..73..959D,2019PhPl...26g2306Y,2020ApJ...898...66Y}.
We thus derive the p.d.f. ${\textsf p}_{\Gamma_l}$ of $\Gamma_l$ as (using $D=3$)~\cite{1991PhRvL..67.2299B}
\begin{align}
{\textsf p}_{\Gamma_l} &\,\sim\, \int {\rm d}\Gamma_{\ell_{\rm c}} \, {\textsf p}_{\Gamma_{\ell_{\rm c}}}\,\int{\rm d}h\, l^{3-d(h)}\,\delta\left[\Gamma_l - \Gamma_{\ell_{\rm c}}\,\left(l/\ell_{\rm c}\right)^h\right]\,.
\label{eq:pdfu}\nonumber\\
&
\end{align}
This p.d.f. remains to be properly normalized. In this formulation, the gradient statistics ${\textsf p}_{\Gamma_l}$ on all scales reduce to a function of the main quantities $v_{\rm A}$, $\ell_{\rm c}$ and the few parameters characterizing $d(h)$, which themselves depend on the physical properties of the turbulence. This offers a first-principles connection between the fundamental statistics of turbulence intermittency and the physics of particle energization.
\smallskip

\noindent\emph{The transport equation--\,}
The CTRW is exactly equivalent to the following kinetic equation for the volume averaged distribution $n_p(t)=4\pi p^2 f(p,\,t)$, where $f(p,\,t)$ represents the angle-averaged distribution function~\cite{1980PhRvL..44...55K,2006RvGeo..44.2003B}:
\begin{align}
\partial_t\,n_p(t)\,=\,\int_{0}^{+\infty}{\rm d}p'\int_{0}^{t}{\rm d}t'\,
\biggl[&\psi\left(p\vert p';\,t-t'\right) n_{p'}(t') \nonumber\\
& - 
\psi\left(p'\vert p;\,t-t'\right) n_p(t')\biggr]\,.
\label{eq:kin-ME}
\end{align}
The kernel $\psi\left(p\vert p';\,t-t'\right)$ differs from the CTRW jump distribution probability $\phi\left(p\vert p';\,t-t'\right)$ introduced earlier, yet the two are related as follows. Denoting with a tilde symbol the Laplace transform in time, and $\nu$ the Laplace variable conjugate to $t-t'$, 
\begin{equation}
\tilde\psi\left(p\vert p';\,\nu\right)\,=\,\frac{\nu\,\tilde\phi\left(p\vert p';\,\nu\right)}{1-\tilde\phi_{p'}\left(\nu\right)}\,,
\label{eq:kern-rel}
\end{equation}
with the short-hand notation $\tilde\phi_{p'}\left(\nu\right)\equiv\int_{0}^{+\infty}{\rm d}p\, \tilde\phi\left(p\vert p';\,\nu\right)$, the subscript $_{p'}$ emphasizing the dependence on $p'$. As discussed above, we characterize the CTRW with a joint probability distribution of the form:
\begin{equation}
\phi\left(p\vert p';\,t-t'\right)\,=\, \varphi\left(p\vert p'\right)\,
\frac{e^{-(t-t')/t_{p'}}}{t_{p'}}\,,
\label{eq:CTRW-ker0}
\end{equation}
recalling that $t_{p'}\equiv l_{\rm g}(p')/c$. Then,
\begin{equation}
\tilde\psi\left(p\vert p';\,\nu\right)\,=\, \frac{\varphi\left(p\vert p'\right)}{t_{p'}}\,,
\label{eq:CTRW-ker}
\end{equation}
in which case the transport equation becomes local in time~\cite{2006RvGeo..44.2003B}, 
\begin{equation}
\partial_t\, n_p(t)\,=\,\int_{0}^{+\infty}{\rm d}p'\,
\left[\frac{\varphi\left(p\vert p'\right)}{t_{p'}} n_{p'}(t) - 
\frac{\varphi\left(p'\vert p\right)}{t_{p}} n_p(t)\right]\,.
\label{eq:kin-ME-loc}
\end{equation}
It takes the form of a master equation for a Markov process, balancing gains and losses at respective rates $t_{p'}$ and $t_{p}$. As $\varphi\left(p\vert p'\right)$ represents the p.d.f. to jump to $p$ from $p'$ in any given amount of time, it is normalized through $\int_{0}^{+\infty}{\rm d}p\,\varphi\left(p\vert p'\right)=1$. Recalling that $\Delta\ln p$ is distributed as $\Gamma_{l_{\rm g}}l_{\rm g}/c$, the p.d.f. $\varphi(p\vert p')$ derives from ${\textsf p}_{\Gamma_l}$ through 
\begin{equation}  
\varphi\left(p\vert p'\right) = \frac{1}{p\, t_{p'}}{\textsf p}_{\Gamma_{l_{\rm g}}}\,,
\label{eq:pdfrel}
\end{equation}
at $\Gamma_{l_{\rm g}} = \ln(p/p')/t_{p'}$, with $l_{\rm g}=l_{\rm g}(p')$.
\smallskip

\noindent\emph{Test against numerical experiments--\,}
\begin{figure}
\includegraphics[width=0.42\textwidth]{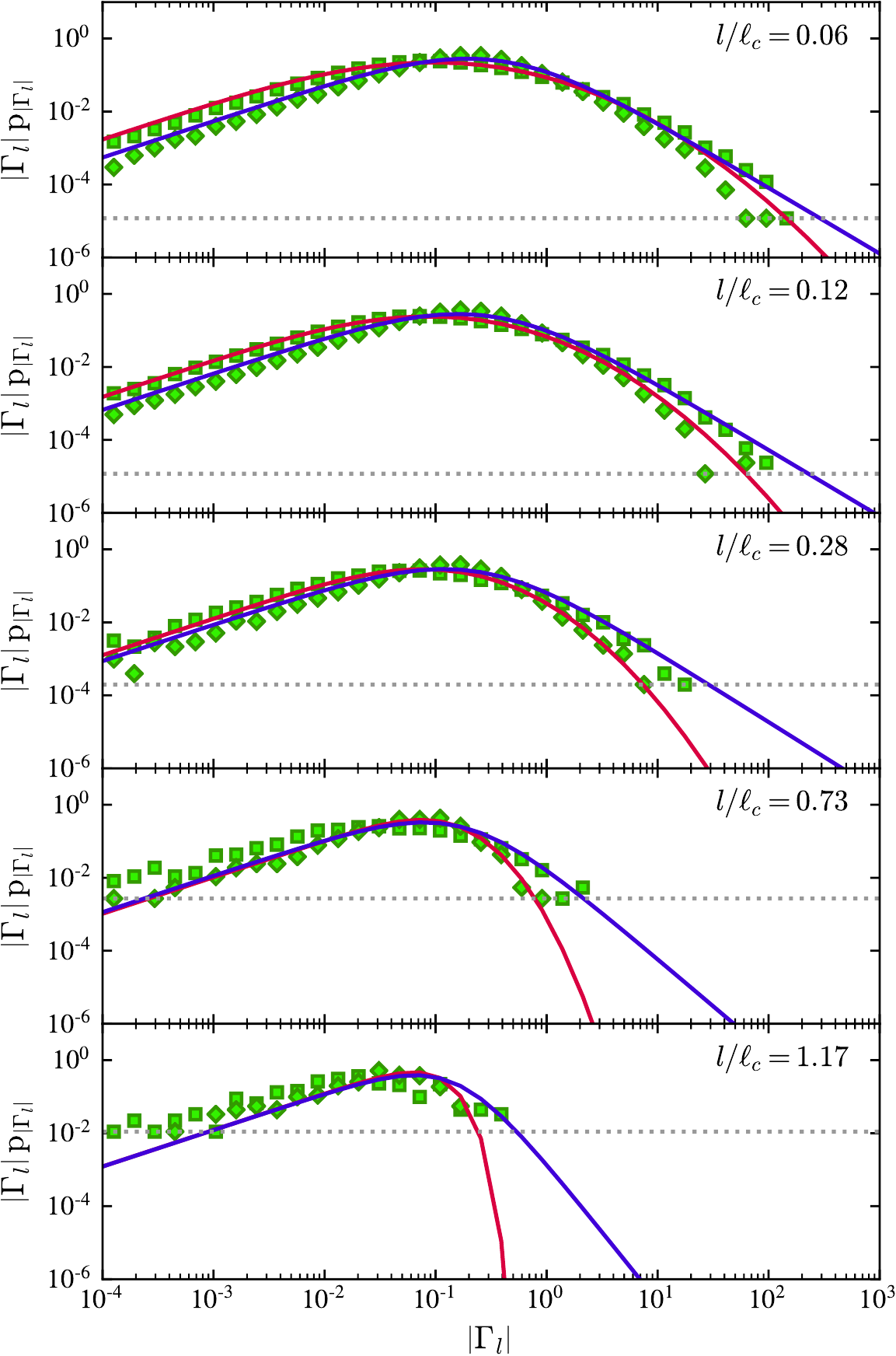}
\caption{Statistics $\vert\Gamma_l\vert \,{\textsf p}_{\vert\Gamma_l\vert}$ of the absolute values of the gradients $\Gamma_l$ ($\Gamma_l$ expressed in units of $c/\ell_{\rm c}$). Symbols: values recorded in a $1\,024^3$ MHD simulation at various coarse-graining scales $l$, as indicated; squares (resp. diamonds): gradients measured along (resp. perpendicular to) the mean magnetic field direction as coarse-grained on scale $l$. The p.d.f. reveals powerlaw tails at large values of $\vert\Gamma_l\vert$ on small scales. Solid red line: adjustment of a multi-fractal lognormal model; solid blue line: a broken powerlaw approximation; dotted line: shot noise level associated to the finite sampling variance.
\label{fig:MFstat}
}
\end{figure}
The above model is now tested on a direct numerical simulation of driven incompressible MHD turbulence ($1\,024^3$ with $1\,024$ time steps)~\cite{2008JTurb...9...31L,2013Natur.497..466E}. This simulation has no guide field, and its units have been set to obtain an Alfv\'en velocity $v_{\rm A}=0.4\,c$; this offers a reasonable compromise between the limits of applicability of this (sub-relativistic) simulation and the value of $v_{\rm A}$ needed to observe acceleration within the duration of the simulation ($2.8\,\ell_{\rm c}/c$); see [Supp.~Mat.~B] for the methods used to extract numerical data from this simulation. 

The statistics of the absolute values of the velocity gradients $\Gamma_l$, parallel and perpendicular to the mean magnetic field and coarse-grained on scale $l$, are shown in Fig.~\ref{fig:MFstat} for various values of $l$. The red solid line represents the adjustment obtained for a log-normal spectrum $d(h) = 3 - \left(h-h_{\textsc{mf}}\right)^2/\left(2\sigma_{\textsc{mf}}^2\right)$, parameterized by $h_{\textsc{mf}}=-0.2$ and $\sigma_{\textsc{mf}}=0.9$, see [Supp.~Mat.~B]. As the random walk has been simplified to one aggregate force term $\Gamma_l$, we have chosen to tune those parameters to provide a fair reconstruction of the ensemble of parallel and perpendicular gradients, rather than fitting one or the other. The solid blue line shows an adjustment of ${\textsf p}_{\vert \Gamma_l\vert}$ by a broken powerlaw approximation, which has the advantage of speeding up the numerical integration of the kinetic transport equation. It takes the form
\begin{equation}
{\textsf p}_{\Gamma_l} \,\sim\, \left[ 1 + \left(\frac{\Gamma_l}{\sigma_{\textsc{bp}}(l)}\right)^{k_0(l)/k_1}\right]^{-k_1}\,,
\label{eq:BPapp}
\end{equation}
with $\sigma_{\textsc{bp}}(l)$ and $k_0(l)$ $l-$dependent quantities; $\sigma_{\textsc{bp}}(l)$ characterizes the width of the distribution before it turns over into the powerlaw behavior with index $\simeq -k_0(l)$, while $k_1= 3$ ensures the smoothness of that transition from core to wing, see  [Supp.~Mat.~B] for details and methodology. Both models use a width $\sigma \simeq 0.3\, v_{\rm A}/\ell_{\rm c}$, with $\sigma = \sigma_{\rm c}$ for $\Gamma_{\ell_{\rm c}}$ [resp. $\sigma=\sigma_{\textsc{bp}}(\ell_{\rm c})$] for the multifractal (resp. broken powerlaw) model. 

From ${\textsf p}_{\Gamma_l}$, we derive $\varphi\left(p\vert p'\right)$ using Eq.~(\ref{eq:pdfrel}) then integrate the transport equation Eq.~(\ref{eq:kin-ME-loc}) to compare the theoretical spectra $n_p(t)$ with experimental ones obtained by tracking a large number of particles in the MHD simulation. We remark here that, first, those two models for ${\textsf p}_{\Gamma_l}$ provide slightly different fits to the measured gradient statistics, therefore the comparison of the corresponding $n_p(t)$ provides a first glance at how the choice of parameters affects those spectra. Second, ${\textsf p}_{\Gamma_l}$ extends to values of $\Gamma_l$ beyond the range where it can be measured in the simulation, allowing in particular for unbounded gains in momentum over a finite timescale $l_{\rm g}/c$; to regularize this, we multiply $\varphi\left(p\vert p'\right)$ by a cut-off $\exp\left[-\left(\ln p-\ln p'\right)^2\right]$, which bounds the maximum gain to the order of unity on that timescale. This theoretical maximum is well motivated here, since $v_{\rm A}$ is not small compared to $c$~\cite{PhysRevD.104.063020}; different choices are possible, but the overall influence does not exceed that associated to the uncertainty affecting ${\textsf p}_{\Gamma_l}$. Third, the model predicts the evolution of momentum in the frame \Rco, hence the comparison to the experimental $n_p(t)$ requires a boost to the simulation frame, which slightly broadens the particle distribution. To minimize this effect, we inject particles at a given momentum $p_0$ in the \Rco\ frame at time $t=0$, integrate the equation then boost the theoretical spectrum to the simulation frame. We thus effectively measure a Green function convoluted with this boost. Finally, the numerical distributions of the velocity gradients have small yet non-zero mean values, implying a slight advection drift toward increasing momenta; it has been taken into account in adjusting ${\textsf p}_{\Gamma_l}$ to the data. Further details on these procedures are provided in [Supp.~Mat.B].

\begin{figure}
\includegraphics[width=0.42\textwidth]{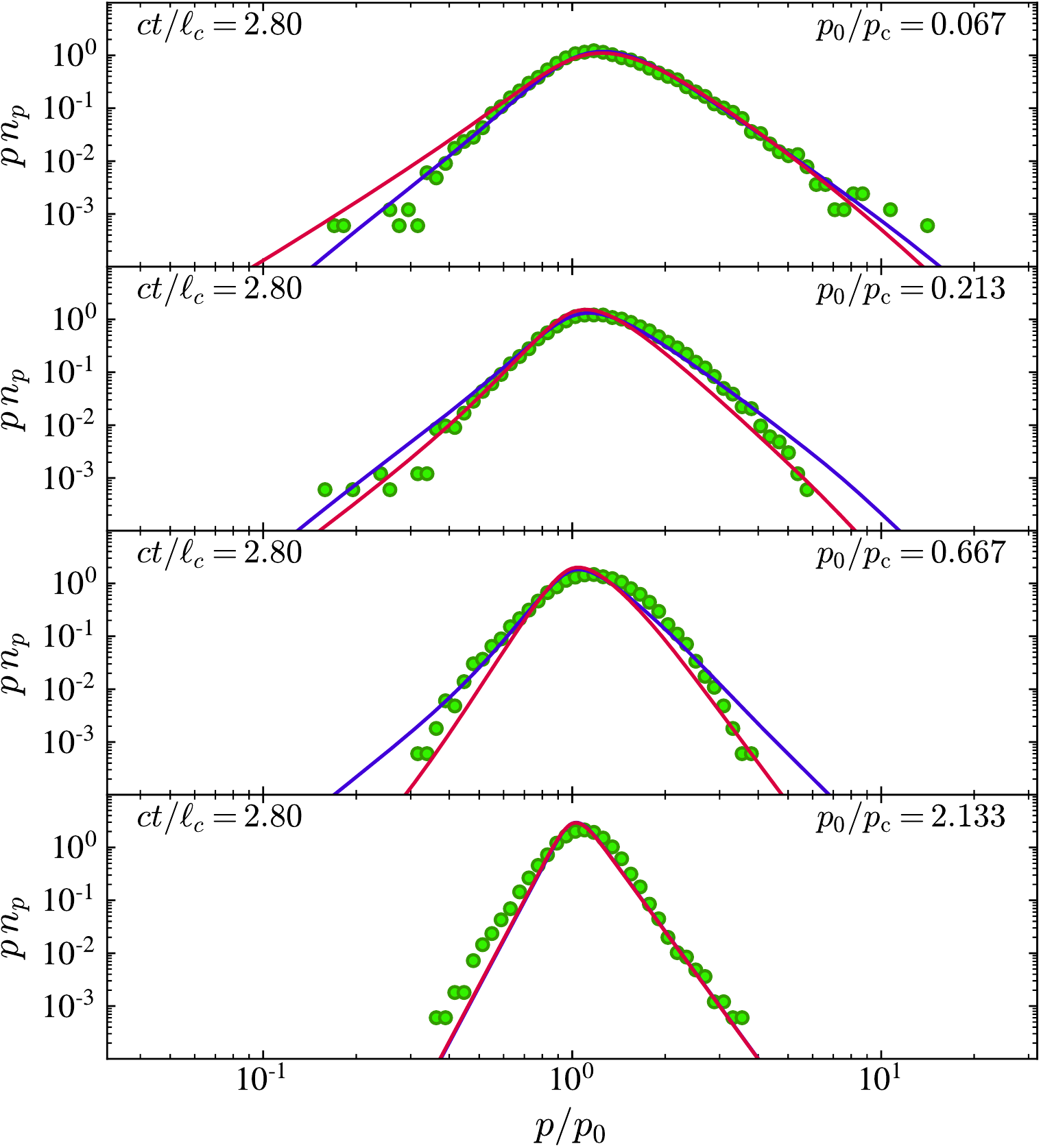}
\caption{Particle spectra obtained at the final time $t=2.8\ell_{\rm c}/c$ for various initial momenta as indicated in units of $p_{\rm c}$; $p_{\rm c}$ is such that $l_{\rm g}(p_{\rm c})=\ell_{\rm c}$. The red and blue lines correspond to the two models shown in Fig.~\ref{fig:MFstat}.
\label{fig:spec_p}}
\end{figure}

\begin{figure}
\includegraphics[width=0.42\textwidth]{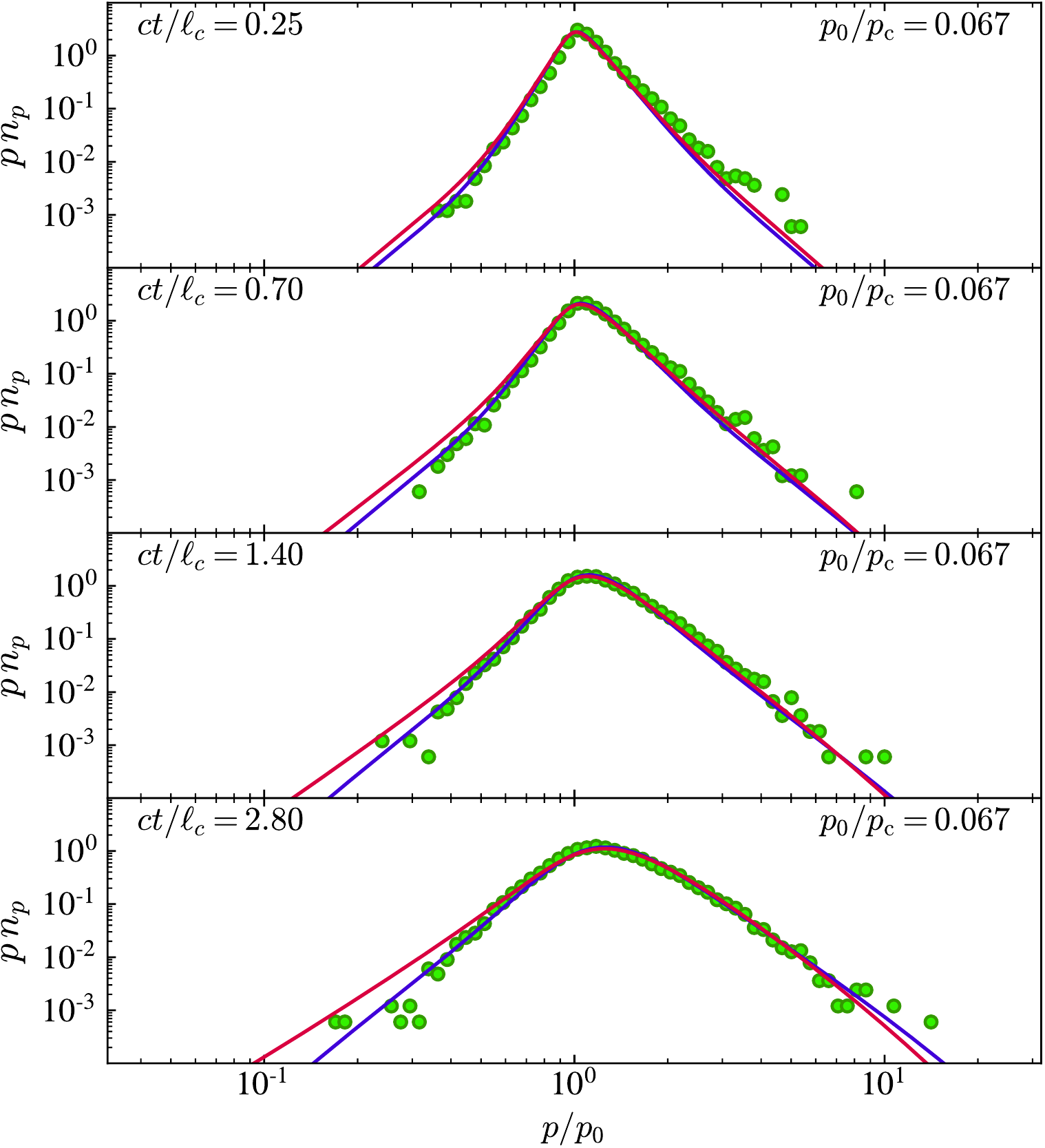}
\caption{Same as Fig.~\ref{fig:spec_p}, now showing the particle spectra at various times, for the smallest initial injection momentum, as indicated.
\label{fig:spec_t}}
\end{figure}

Those theoretical models (red and blue curves, following the conventions of Fig.~\ref{fig:MFstat}) are compared in Figs.~\ref{fig:spec_p} and \ref{fig:spec_t} to the momentum- and time-dependent Green functions obtained by tracking a large number of test particles in the  simulation cube. The test particles have been propagated in the time-dependent snapshots of the simulation, {\it i.e.} the time evolution of the electromagnetic fields has been properly taken into account. In Fig.~\ref{fig:spec_p}, the spectra are plotted {\it vs} $p/p_0$ for increasing values of the initial momentum $p_0$, from $p_0/p_{\rm c}=0.067$ to $p_0/p_{\rm c}=2.1$, thus covering a dynamic range of $1.5$ decade; $p_{\rm c}$ denotes the momentum such that $l_{\rm g}(p_{\rm c})=\ell_{\rm c}$, {\it i.e.} the coarse-graining becomes comparable to the coherence scale. At $p_0>p_{\rm c}$, the spectrum narrows down: the acceleration decreases because the particle then sees the turbulence as a collection of incoherent cells of small extent relative to its gyroradius. In Fig.~\ref{fig:spec_t}, the spectra are plotted for $p_0/p_{\rm c}=0.067$ up to the maximal integration time of the simulation, $t=2.8\,\ell_{\rm c}/c$. This rather satisfactory comparison between theoretical and experimental Green functions supports the present picture. 
\smallskip

\noindent\emph{Discussion and perspectives--\,}
As in the original Fermi picture, the efficiency of acceleration scales with the turbulent Alfv\'en velocity $v_{\rm A}$~\footnote{In the presence of a mean field $B_0$ (corresponding Alfv\'en velocity $v_{\rm A\,0}$), $v_{\rm A}=(\delta B/B_0)v_{\rm A\,0}$ ($\delta B$ turbulent component), therefore $\delta B/B_0$ impacts acceleration at fixed $v_{\rm A\,0}$.}, {which controls the magnitude of the random force through its influence on $\Gamma_{\ell_{\rm c}}$.  At lower $v_{\rm A}$, one observes a softer spectrum at a given time, but as time passes, the spectrum becomes harder; two different values of $v_{\rm A}$ eventually give comparable spectra at times rescaled by $1/v_{\rm A}^2$.
Figures~\ref{fig:spec_p} and \ref{fig:spec_t}  indicate that powerlaw spectra are generic, as observed in kinetic simulations~\cite{17Zhdankin,18Comisso}. Those powerlaws find their origin in the large excursions that particles momenta can undergo in sparse regions of intense gradients~\cite{PhysRevD.104.063020,2020MNRAS.499.4972L}, here captured by the extended wings of $\varphi\left(p'\vert p\right)$. In the absence of those, a Kramers-Moyal expansion would indeed reduce Eq.~(\ref{eq:kin-ME-loc}) to the Fokker-Planck form with diffusion coefficient $D_{pp}=\int{\rm d}p'\,\left(p'-p\right)^2\,\varphi\left(p'\vert p\right)/t_{p}$ and advection coefficient $A_p=\int{\rm d}p'\,\left(p'-p\right)\,\varphi\left(p'\vert p\right)/t_{p}$. 

The present scenario naturally accounts for the recent observation that some particles can see their energy increase exponentially fast in localized regions, up to an energy gain of a few~\cite{2022ApJ...928...25P}, by virtue of $\dot p=p\,\Gamma_l$ with $\Gamma_l$ varying on scales of extent $l$; see also \cite{PhysRevD.104.063020,2016MNRAS.458.2584B}. As scattering is here dominated by magnetic mirrors, which become intermittent on small scales, one also anticipates anomalous spatial transport. Inspection of numerical data confirms that some particles preserve their pitch-angles over long distances $\gtrsim\ell_{\rm c}$, while others suffer strong deflection on short distances. This might account for spatial superdiffusion events observed in recent simulations~\cite{2022ApJ...926...94M} and trapping in others~\cite{2020ApJ...894..136T}.

The present approach differs from that of Ref.~\cite{2020ApJ...893L...7W}, which extracts from numerical experiments empirical momentum-dependent functions $A_p =\langle \Delta p\rangle/\Delta t$ and $D_{pp}=\langle \Delta p^2\rangle/\Delta t$. It also differs from Ref.~\cite{2017PhRvL.119d5101I}, which models transport in momentum space through a fractional Fokker-Planck equation, describing the random walk as a L\'evy process. The present $\varphi(p'\vert p)$ is more akin to a truncated L\'evy flight, which contains flat tails but well-defined high-order moments. This implies that, at asymptotic times, the random walk will behave as some Brownian motion, due to central-limit convergence; however, the extended wings of $\varphi\left(p'\vert p\right)$ slow down this convergence quite appreciably, as expressed by the Berry-Ess\'een theorem~\cite{1995PhRvL..74.4959S}, hence, for practical matters, the present non-Fokker-Planck form remains required. This transport equation remains amenable to extensions: for instance, wave-particle resonant interactions, if effective~\cite{2000PhRvL..85.4656C,2002PhRvL..89B1102Y}, could be included by adding an extra diffusion term; similarly for particle heating in small-scale non-ideal electric fields, which ensures injection into the Fermi process in kinetic simulations~\cite{17Zhdankin,18Comisso,2019ApJ...886..122C,2020ApJ...893L...7W,2021ApJ...922..172Z}; finally, including standard radiative and escape terms would allow to model the emerging spectra from astrophysical sources. 

In summary, the present work has established a novel, general framework for implementing stochastic Fermi acceleration in a realistic, collisionless turbulent bath, opening a  connection between the statistics of the intermittent gradients of the velocity of magnetic field lines and the rate of energization. Improved insight on turbulence intermittency could ultimately allow a first-principles calculation of particle spectra. The applicability of this formalism thus extends beyond that of the numerical simulation of incompressible MHD turbulence against which it was successfully tested. 
\smallskip

\noindent \emph{Acknowledgments:} This work has been supported by the Sorbonne Universit\'e DIWINE Emergence-2019 program and by the ANR (UnRIP project, Grant No.~ANR-20-CE30-0030). The possibility to use the resources of the JH Turbulence Database (JHTDB), which is developed as an open resource by the Johns Hopkins University, under the sponsorship of the National Science Foundation, is gratefully acknowledged.

\bibliography{refs}

\pagebreak
\widetext
\begin{center}
\textbf{Supplemental Material to ``First-principles Fermi acceleration in magnetized turbulence''}
\end{center}
\setcounter{equation}{0}
\setcounter{figure}{0}
\setcounter{table}{0}
\setcounter{page}{1}
\makeatletter
\renewcommand{\theequation}{A\arabic{equation}}
\renewcommand{\thefigure}{A\arabic{figure}}

\appendix

\renewcommand{\theequation}{A\arabic{equation}}
\renewcommand{\thefigure}{A\arabic{figure}}

\section{Methods -- stochastic model}\label{sec:appstoch}
The random walk is written: $\dot p = p\,\Gamma_{l_{\rm g}}$, where $\Gamma_{l_{\rm g}}$ is expressed as a gradient of the velocity field $\boldsymbol{v_E}$ coarse-grained on scale $l$ [Eq.~(1)]. If $\Gamma_{l_{\rm g}}$ were Gaussian distributed, with a vanishing coherence time, meaning a p.d.f. characterized by $\langle\Gamma_{l_{\rm g}}\rangle$ (drift) and $\left\langle\delta\Gamma_{l_{\rm g}}(t)\delta\Gamma_{l_{\rm g}}(t')\right\rangle \propto \delta(t-t')$ ($\delta\Gamma_{l_{\rm g}}\equiv\Gamma_{l_{\rm g}}-\langle\Gamma_{l_{\rm g}}\rangle$)~\cite{1989fpem.book.....R}, the random walk would be described as Brownian motion and the transport equation would take a Fokker-Planck form.  This is not the case here, since the random process describes the influence of velocity structures whose scales $l$ enter the range $l_{\rm g} < l < \ell_{\rm c}$, and whose amplitudes are not Gaussian distributed. One must thus consider the full transition kernel $\phi\left(p\vert p';\,t-t'\right)$ that defines the probability to transit from momentum $p'$ to $p$, from time $t'$ to $t$. The stochastic process then takes the form of a continuous-time random walk: the time step between two jumps becomes random distributed and its distribution depends on the location in momentum space. This description fits well the present picture, because the coarse-graining scale $l_{\rm g}(p')$ depends on momentum. In order to simplify the description, and to bring the transport equation to a tractable form, two assumptions are made: (1) structures on the coarse-graining scale $l_{\rm g}$ dominate the acceleration process; (2) the p.d.f. of the time interval $\Delta t$ between two momentum jumps is exponentially distributed, meaning ${\rm p.d.f.}(\Delta t)=\exp(-\Delta t/\tau)/\tau$, with $\tau = t_{p'} \equiv l_{\rm g}/c$ by application of (1) (for particles moving at $v\sim c$). 

Assumption (2) is weak, in the sense that a Monte Carlo implementation of the above stochastic process indicates that other p.d.f. for the transition time, with equal mean waiting time, lead to nearly identical particle spectra. The choice of an exponential p.d.f. simplifies however the derivation of the transport equation.

Assumption (1) rests on the premise that the stochastic force increases in amplitude as one goes to smaller scales, and so does intermittency. This also agrees with the discussion of the comparative influence of various scales $l$ on energization in~\cite{PhysRevD.104.063020}. In order to test this assumption, the transport equation has been integrated to obtain particle spectra for different choices of the dominant scale $l$. Namely, the mean waiting time $\tau$ between two jumps has been set to $\tau={\rm max}\left(l/c,\,l_{\rm g}/c\right)$, while $\Delta \ln p$ has been assumed to scale as $\Gamma_{c\,\tau} \, \tau$ as before. Setting $l = l_{\rm g}$ for all $p'$, one recovers $\tau=t_{p'}$ and the transport equation discussed in the main text; by contrast, fixing $l$ to some constant value mimics a situation in which structures on scale $l$ dominate the physics of acceleration. This exercise reveals that the spectra become steeper for $l\gtrsim 0.1\ell_{\rm c}$, indicating that the largest scales provide a sub-dominant contribution to acceleration. This falls in line with the above expectations: as intermittency becomes weak on large scales, the development of powerlaw tails is suppressed. For the values of $p$ considered here in the comparison to numerical data, {\it i.e.} $l_{\rm g}(p_0)/\ell_{\rm c}\gtrsim0.07$, this observation supports assumption (1).  

For momenta $p\geq p_{\rm c}$, such that $l_{\rm g}\geq l_{\rm c}$, the mean waiting time must be kept constant at $\ell_{\rm c}/c$, as this properly describes the coherence time of the random force that is exerted on the particle. On such timescale, the particles suffers energy gain or loss of the order of $\vert\Delta \ln p\vert\simeq (v_{\rm A}/c) \ell_{\rm c}/l_{\rm g} \sim \Gamma_{\ell_{\rm c}}\,\ell_{\rm c}^2/l_{\rm g}c$, the additional factor $\ell_{\rm c}/l_{\rm g}$ accounting for the fact that those particles decouple from the turbulence due to their large magnetic rigidity.

Finally, the stochastic process can be generalized to sub-relativistic particles ($v\ll c$) through the direct substitution $t_p \rightarrow l_{\rm g}/v$ for the mean waiting time, in which case $\Delta \ln p$ is distributed as $\Gamma_{l_{\rm g}} l_{\rm g}/v$. The definition of $\Gamma_{\rm acc}$ indicates that it scales as $v_{\rm A}/v$ relatively to the other two; hence, at $v\ll v_{\rm A}$, this term should dominate the energization. As the intermittent statistics of $\Gamma_{\rm acc}$ remain comparable to those of $\Gamma_\parallel$ and $\Gamma_\perp$, this should not affect the particle spectra significantly.  

\renewcommand{\theequation}{B\arabic{equation}}
\renewcommand{\thefigure}{B\arabic{figure}}

\section{Methods -- data extraction and manipulation out of a $1024^3$ MHD numerical simulation}\label{sec:appnum}
The numerical data is extracted from a direct numerical simulation of incompressible MHD on a periodic grid containing $1\,024^3$ cells, sampled on $1\,024$ time steps. 
 This simulation, which is made available for public use from the Johns Hopkins University Turbulence Database, available from:\\ 
\href{http://turbulence.pha.jhu.edu/Forced_MHD_turbulence.aspx}
{http://turbulence.pha.jhu.edu/Forced\_MHD\_turbulence.aspx.}~\cite{2013Natur.497..466E}, is visco-resistive, with magnetic Reynolds number at the Taylor scale $\simeq140$, and magnetic Prandtl number unity. The $1\,024$ cube snapshots have been archived after the simulation has reached a statistically stationary state; the properties of the turbulence evolve little over those additional $1\,024$ time steps. The turbulence is driven through an external random force acting on velocity fluctuations on a stirring scale equal to $0.5$ in units of the size of the simulation box. The corresponding correlation length for the magnetic field, as defined in the database, is $\simeq 0.1$ in those units. In the present work, we adopt $\ell_{\rm c}=0.14$ for reasons detailed below. 

There is no guide field, hence the Alfv\'en velocity $v_{\rm A}$ denotes that of turbulent fluctuations. The units of the simulation are fixed to obtain $v_{\rm A}=0.41\,c$; accordingly, the rms velocity (as measured without coarse-graining) is $\langle\delta u^2\rangle^{1/2}\simeq 0.4\,c$. This choice of units allows to remain within the limits of applicability of the simulation, which is not special-relativistic, and at the same time to offer the possibility to observe particle acceleration on the finite timescale of the simulation ($2.8\,\ell_{\rm c}/c$). A fraction $0.1\,$\% of the volume contains velocities in excess of $0.8\,c$, and $0.03\,$\% of the volume in excess of $0.88\,$c, therefore this compromise is reasonable. See also further below.

\subsection{Statistics of velocity gradients}
The statistics of velocity jumps are obtained through Monte Carlo sampling of the simulation cube. For a given coarse-graining scale $ l$ (expressed in units of the cube size), $N_{l}\simeq 1/{l}^3$ points are drawn at random through the cube; $N_{l}$ is thus chosen so as to obtain a maximum number of points without oversampling the volume. $N_{ l}$ decreases rapidly with increasing coarse-graining scale $ l$, up to $N_{ l}\sim \mathcal O(300)$ on the coherence scale $ l\simeq \ell_{\rm c}$. Correspondingly, the sample variance $\sim 1/N_{ l}$ becomes non-negligible on those scales.

Using the numerical tools provided by the database, the values of the plasma velocity 
${\boldsymbol{v_{\rm p}}}_l$, the magnetic field $\boldsymbol{B}_l$ and their spatial derivatives $\partial_i {{v_{\rm p}}_j}_l$, $\partial_i {B_j}_l$ are coarse-grained on scale $l$. Those are used compute the velocity field ${\boldsymbol{v_{E}}}_l$ through
\begin{equation}
{\boldsymbol{v_{E}}}_l = {\boldsymbol{v_{\rm p}}}_l - \left({\boldsymbol{v_{\rm p}}}_l\cdot\boldsymbol{b}_l\right)\,\boldsymbol{b}_l\,,
\label{eq:vEvp}
\end{equation}
with $\boldsymbol{b}_l = \boldsymbol{B}_l/\left\vert \boldsymbol{B}_l\right\vert$. This amounts to calculating the electric field using ideal Ohm's law $\boldsymbol{E}_l=-{\boldsymbol{v_{\rm p}}}_l\times \boldsymbol{B}_l/c$ then ${\boldsymbol{v_{E}}}_l=c\,\boldsymbol{E}_l\times \boldsymbol{B}_l/B_l^2$. Resistive corrections to ideal Ohm's law can be neglected on the scales of interest. The spatial derivatives of ${\boldsymbol{v_{E}}}_l$ are similarly derived from those of ${\boldsymbol{v_{\rm p}}}_l$ and $\boldsymbol{B}_l$.

Those field values and their derivatives can be used to reconstruct the quantities entering the force terms, in particular $\Gamma_\parallel$ and $\Gamma_\perp$: keeping track of the scale $l$ over which they are defined through a subscript,  
\begin{align}
{\Gamma_\parallel}_l &\,=\, -\boldsymbol{b}_l\cdot\left(\boldsymbol{b}_l\cdot\boldsymbol{\nabla}\right)\,{\boldsymbol{v_{E}}}_l\,,\nonumber\\
{\Gamma_\perp}_l &\,=\, -\frac{1}{2}\left[\boldsymbol{\nabla}\cdot{\boldsymbol{v_{E}}}_l-\boldsymbol{b}_l\cdot\left(\boldsymbol{b}_l\cdot\boldsymbol{\nabla}\right)\,{\boldsymbol{v_{E}}}_l\right]
\label{eq:gammacg}
\end{align} 

\subsection{Multifractal model}
Those gradients ${\Gamma_\parallel}_l$ and ${\Gamma_\perp}_l$ are random variables that become increasingly non-Gaussian at small scales $l\ll\ell_{\rm c}$. Such statistics are conveniently discussed within the multifractal picture of turbulence intermittency~\cite{Frisch..Parisi.85}, which ascribes a local scaling exponent $h(\boldsymbol{x})$ to a position $\boldsymbol{x}$, and which describes the set of locations $\boldsymbol{x}$ with index $h(\boldsymbol{x})$ as a fractal of dimension $d(h)$. The spectrum $d(h)$ is a continuous function of $h$. 

Intermittency can also be captured through the structure functions, {\it e.g.} $S_n(l) = \langle \vert \Gamma_l \vert^n\rangle$, and their scaling exponents $S_n(l) \propto l^{\zeta_n}$. The scaling exponent $\zeta_n$ and the spectrum $d(h)$ are however in one-to-one correspondence via the Legendre transform $\zeta_n={\rm min}_{h}\left[h\,n + D - d(h)\right]$~\cite{Frisch..Parisi.85} ($D$ number of spatial dimensions). Therefore, using one or the other is a matter of choice. We rely here on $d(h)$ to match the statistics of $\Gamma_l$ on scale $l$, in order to characterize the probability distribution function (p.d.f.) of momentum jumps. 

This spectrum $d(h)$ is generally described by a log-normal form~\cite{2019JFM...867P...1D},
\begin{equation}
d(h) = D_0 - \frac{\left(h-h_{\textsc{mf}}\right)^2}{2\sigma_{\textsc{mf}}^2}\,,
\label{eq:lognorm}
\end{equation}
 or a log-Poisson model~\cite{1994PhRvL..72..336S,1994PhRvL..73..959D,2019JFM...867P...1D}
\begin{equation}
d(h) = D_0 - \frac{(h-h_0)}{\ln \beta}\left[1-\ln\left(\frac{h-h_0}{(D-D_0)\vert\ln\beta\vert}\right)\right]\,.
\label{eq:logpois}
\end{equation}
In these formulations, $D_0$ represent the dimension of the most intermittent structures, while $h_\textsc{mf}$, $\sigma_\textsc{mf}$, $h_0$ and $\beta$ are parameters. Currently, those parameters are adjusted to simulation data, but ultimately, one may hope to gain fundamental insight on their values.

Theories of turbulence intermittency commonly discuss the statistics of the Els\"asser fields $\boldsymbol{\delta {z_\pm}_l} = \boldsymbol{\delta v_l} \pm \boldsymbol{\delta b_l}$, expressed in terms of the plasma velocity fluctuation $\boldsymbol{\delta v_l} = {\boldsymbol{v_{\rm p}}}_l\left(\boldsymbol{x}+\boldsymbol{l}\right) - {\boldsymbol{v_{\rm p}}}_l\left(\boldsymbol{x}\right)$ coarse-grained on scale $l$ and the magnetic field fluctuation (normalized to Alfv\'en velocity units) 
$\boldsymbol{\delta b_l} = \left[\boldsymbol{B}_l\left(\boldsymbol{x}+\boldsymbol{l}\right) - \boldsymbol{B}_l\left(\boldsymbol{x}\right)\right]/\sqrt{4\pi\rho}$ ($\rho$ plasma mass density). The quantities $\delta z_\pm\equiv\boldsymbol{\delta {z_\pm}_l}\cdot\boldsymbol{l}/\vert\boldsymbol{l}\vert$ are well described by log-Poisson models with {\it e.g.}, $D_0=2$, $h_0=\frac{1}{9}$ and $\beta \simeq 0.69$~\cite{2000PhPl....7.4889B,2003PhRvE..67f6302M,2004PhRvL..92s1102P,2015ApJ...807...39C}. Such a model reproduces satisfactorily the p.d.f. of $\delta z_\pm$ in the present MHD simulation, as explicitly verified.

The measured statistics of ${\Gamma_\parallel}_l$ and ${\Gamma_\perp}_l$ differ from those of $\delta z_\pm/\vert\boldsymbol{l}\vert$ (the additional $1/\vert\boldsymbol{l}\vert$ is introduced to convert the original Els\"asser quantity into a gradient). The p.d.f. of ${\Gamma_\parallel}_l$ and ${\Gamma_\perp}_l$ are systematically more extended and they display harder powerlaws at large values of the arguments. This difference can be attributed to the following: ${\Gamma_\parallel}_l$ and ${\Gamma_\parallel}_l$ are defined in terms of the velocity field $\boldsymbol{v_E}$, whereas $\delta z_\pm$ is related to the plasma velocity $\boldsymbol{v_{\rm p}}$; however, gradients of those velocity fields differ by quantities related to the curvature of magnetic field lines~\cite{PhysRevD.104.063020}, while the statistics of the field line curvature displays hard powerlaw tails with slope close to $-2.5$~\cite{2019PhPl...26g2306Y,2020ApJ...898...66Y}. In the following, we use a simple log-normal model, which turns out to provide a satisfactory description of their p.d.f.; more work is warranted to relate the statistics of ${\Gamma_\parallel}_l$ and ${\Gamma_\perp}_l$ with those of $\delta z_\pm$.

\subsection{Multi-fractal fitting procedure}
The multi-fractal model for ${\Gamma_\parallel}_l$ and ${\Gamma_\perp}_l$ is constructed as described in the text. On the coherence scale $\ell_{\rm c}$ of the turbulence, Gaussian statistics are well motivated to model $p_{\Gamma_{\ell_{\rm c}}}$ while on smaller length scales, the statistics of velocity gradients become increasingly non-Gaussian. We have set $\ell_{\rm c}=0.14$ in units of the cube size, rather than $0.1$ as it is defined in the database, because non-Gaussian wings remain visible on the latter scale and start to disappear on the former. This rescaling exerts no significant influence on the comparison of the theoretical Green functions to the data.

The multi-fractal model corresponds to the log-normal spectrum $d(h)= 3 - \left(h-h_{\textsc{mf}}\right)^2/\left(2\sigma_{\textsc{mf}}^2\right)$, characterized by  $\sigma_{\textsc{mf}}=0.9$ and $h_{\textsc{mf}}=-0.2$. 

The distributions $p_{\Gamma_{l}}$ (with $\Gamma_l$ denoting generically ${\Gamma_\parallel}_l$ or ${\Gamma_\perp}_l$) are slightly asymmetric around zero. This asymmetry appears to depend on scale, it differs between the parallel and perpendicular direction and it is not always well resolved out of the noise of the numerical data. Yet, it implies a non-vanishing mean value for $\Gamma_l$ and therefore net advection in momentum space. This mean value can be measured by averaging over the simulation cube; it is discussed in Ref.~\cite{PhysRevD.104.063020}. Net advection can also be measured in the spectra obtained by particle tracking in the MHD simulation (discussed next). To model this asymmetry while retaining a minimal number of parameters, we proceed as follows. We first break the distribution $p_{\Gamma_l}$ over positive and negative values of $\Gamma_l$. Both parts are characterized by the same parameters except the width $\sigma_{\rm c}$ characterizing the fluctuations on the integral scale, which takes value $\sigma^+_{\rm c}\simeq 0.10\,c/\ell_{\rm c}$ for positive values of $\Gamma_l$, and $\sigma^-_{\rm c}\simeq 0.07\,c/\ell_{\rm c}$ for negative values of $\Gamma_l$.  

The distribution shown in the main text plots the distribution for the absolute values $\vert\Gamma_l\vert$, thus summing those two partial distributions. Overall, the distributions of $\Gamma_l$ on all scales are characterized by four parameters, whose uncertainty is of the order of $0.1$ for $\sigma_{\textsc{MF}}$ and $h_{\textsc{MF}}$, and $0.05\,c/\ell_{\rm c}$ for what regards $\sigma^\pm_{\rm c}$. This uncertainty exceeds slightly the difference between $\sigma^+_{\rm c}$ and $\sigma^-_{\rm c}$: their difference has been adjusted through its influence over the advection rate, as measured from the final particle spectra.

A drawback of the above multifractal model is that the integral that defines $p_{\Gamma_l}$ cannot be expressed analytically. To speed up computations, it proves useful to introduce an approximation of the $\Gamma_l$ statistics, modeled as the broken powerlaw described in the main text. The parameters used to adjust the gradient statistics are: $\sigma_{\textsc{bp}}(l) = \sigma_{\textsc{bp}}(\ell_{\rm c}) \left(l/\ell_{\rm c}\right)^{-0.3}$, $k_0(l) = 2.8 + 2\left(l/\ell_{\rm c}\right)^2$ and $k_1=3$. The dependence of $k_0$ on $l$ is to guarantee that, as $l\rightarrow\ell_{\rm c}$, where the distribution should become approximately Gaussian, the powerlaw wing does not play a role anymore. 

For the same reasons as above, we break this distribution into two sub-distributions, one for positive values of $\Gamma_l$, one for negative values. They are characterized by the same parameters, except $\sigma_{\textsc{bp}}(\ell_{\rm c})$, which breaks into $\sigma^+_{\textsc{bp}}(\ell_{\rm c})=0.17\,c/\ell_{\rm c}$ and $\sigma^-_{\textsc{bp}}(\ell_{\rm c})=0.13\,c/\ell_{\rm c}$. The uncertainties on those parameters are estimated to be $\simeq 0.05\,c/\ell_{\rm c}$ for $\sigma_{\textsc{bp}}(\ell_{\rm c})$, and $\simeq 0.2$ for $k_0(l)$. 

Finally, it should be stressed that better fits could be obtained by allowing for more degrees of freedom, but the choice has been made here to retain a maximum of simplicity in view of the uncertainty characterizing the numerical data and the number of simplifications made in building up the theoretical model.

\subsection{Particle tracking}
To reconstruct the Green functions describing the evolution of the momentum distribution function in time, a large number of particles are tracked in the simulation (typically $24\,000$ per value of the initial momentum). Each test particle is initialized at time $t=0$ at a random position, with a random orientation in velocity space and with a given momentum $p_0$ in the \Rco\ frame in which the electric field vanishes. 

The particle position and momentum is advanced using a Boris pusher. At each time step of duration $0.1\,r_{\rm g}(p_0)/c$, $r_{\rm g}(p_0)$ denoting the initial gyroradius $p_0c/eB$, the pusher queries the database to retrieve the values of $\boldsymbol{v_{\rm p}}$ and $\boldsymbol{B}$ at the position of the particle and at the corresponding time. The time evolution of the electromagnetic fields is thus properly taken into account. Those values are obtained using high-order Lagrangian interpolation. The field values are not coarse-grained here, hence the code advances the particles using the true  time-dependent Lorentz force. The electric field is reconstructed using ideal Ohm's law $\boldsymbol{E}=-\boldsymbol{v_{\rm p}}\times\boldsymbol{B}/c$. Finally, the Green functions are reconstructed by calculating the momentum distribution function of particles, as a function of time $t$ and as a function of initial momentum $p_0$. 

In order to compare the predictions of the theoretical transport equation with the numerical data, the theoretical spectra, defined in \Rco, are boosted to the simulation frame. Noting that the boost multiplies the energies in \Rco\ by a factor $(1 + \mu\,v_E)/\sqrt{1-v_E^2/c^2}$, with $\mu$ the cosine of the angle between the particle direction and the velocity field $\boldsymbol{v_E}$ at that location ($v_E=\vert\boldsymbol v_E\vert$), we draw a large number of values $\mu$ uniformly distributed in $[-1,1]$ and we extract a large number of velocity values $v_E$ at random points in the simulation to construct that kernel. The MHD simulation is not special-relativistic, hence values of $v_E$ in excess of $0.87c$ (corresponding to Lorentz factor $2$) are discarded; this concerns however a tiny fraction of the numerical grid ($\lesssim 0.03\,\%$).

We consider values of $p_0$ such that the wavenumber scale $k\sim r_{\rm g}^{-1}$ associated to the initial gyroradius $r_{\rm g}(p_0)$ falls within the inertial range of the turbulence cascade, above the dissipative range. This corresponds, in practice to values $10^{-2} \,\ell_{\rm c}\lesssim r_{\rm g}(p_0)\lesssim 0.15 \,\ell_{\rm c}$. In terms of coarse-graining scales, which are used in the theoretical computation, this corresponds to $0.06\lesssim l_{\rm g}/\ell_{\rm c}\lesssim 1$.

\subsection{Sensitivity to the fitting procedure}
\begin{figure}
\includegraphics[width=0.45\textwidth]{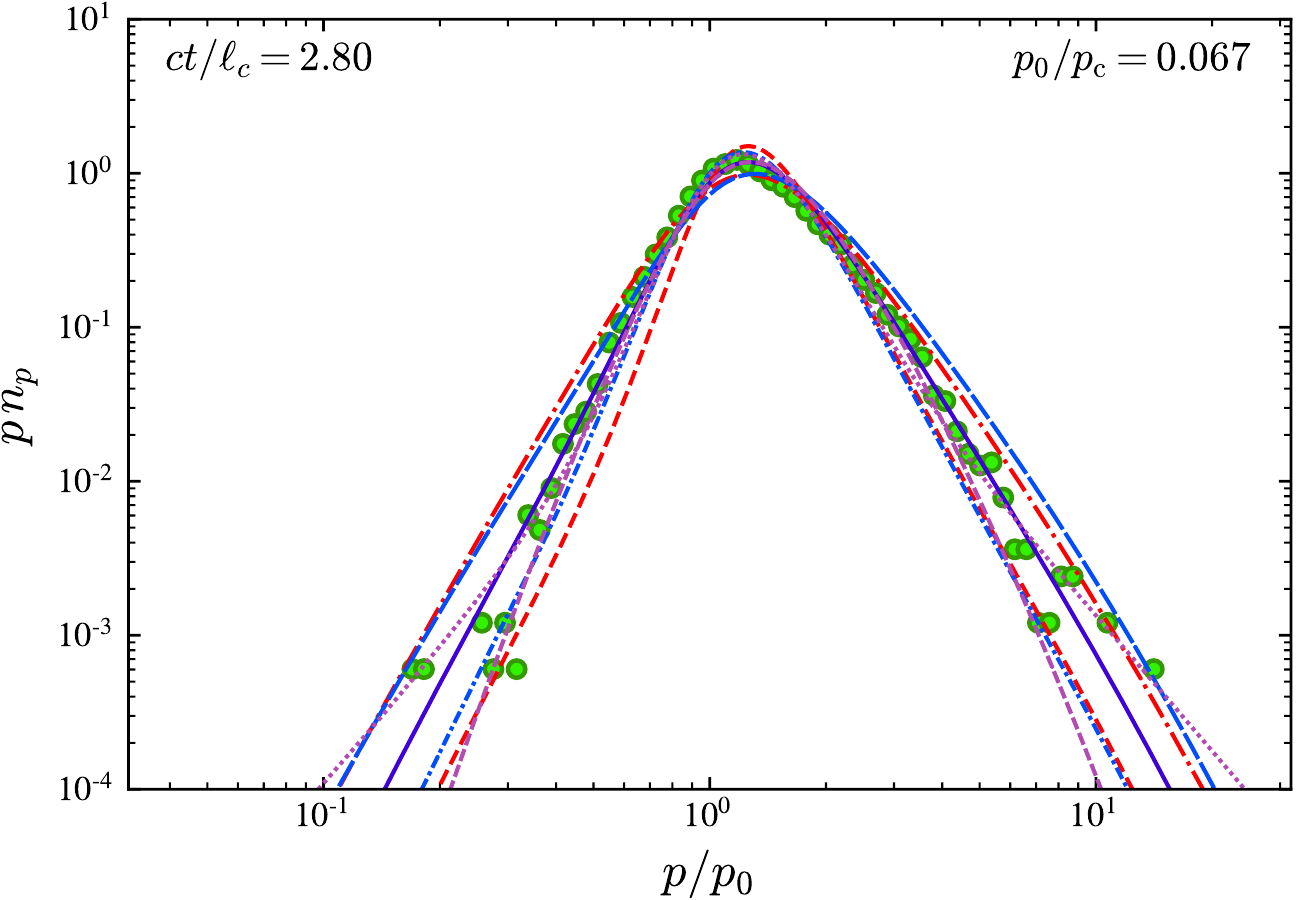}
\caption{Particle spectra $p\,n_p(t)$ obtained by tracking a large number of particles in the MHD simulation (symbols), with theoretical spectra overlaid. Solid blue line: spectrum for the broken powerlaw model with parameters adopted in the main text; dashed (resp. dash-dotted) red lines: spectra obtained by decreasing (resp. increasing) $\sigma_{\textsc{bp}}$ by $0.05$; long-dashed (resp.  short-dash-dotted) blue lines: spectra obtained for $k_0(l)= 2.6 + 2\left(l/\ell_{\rm c}\right)^2$ [resp. $k_0(l)= 3.0 + 2\left(l/\ell_{\rm c}\right)^2$]; densely dashed magenta line: changing the cut-off function $\exp\left[-\left(\ln p-\ln p'\right)^2/a\right]$ from $a=1$ (main text) to $a=0.3$; densely dotted magenta line: same, but with $a=3$.
See text for details.
\label{fig:var}
}
\end{figure}
The spectra $n_p(t)$ that are derived by integrating the kinetic equation depend directly on the model of velocity gradient statistics. The two models plotted in red (multi-fractal) and blue (broken powerlaw) offer a first glance at this influence. Additionally, we plot in Fig.~\ref{fig:var} a number of spectra obtained by varying the parameters of the broken powerlaw approximation within the quoted uncertainties. The solid blue corresponds to the spectrum discussed in the main text, with the choice of parameters indicated above. The dashed and dash-dotted red lines show the spectra obtained by respectively, decreasing or increasing the values of $\sigma_{\textsc{bp}}(\ell_{\rm c})$ by $0.05\,c/\ell_{\rm c}$. The long-dashed and short-dash-dotted blue lines show the spectra obtained by changing the $2.8$ exponent of $k_0$ to, respectively, $2.6$ and $3.0$. Finally, the densely dashed and densely dotted magenta lines show the effect of changing the cut-off function $\exp\left[-\left(\ln p-\ln p'\right)^2/a\right]$ from $a=1$ (main text) to respectively, $a=0.3$ and $a=3$. The model for $a=0.3$ also assumes: $k_0(l) = 2.6 + 2(l/\ell_{\rm c})^2$, $\sigma_{\textsc{bp}}^-=0.15\,c/\ell_{\rm c}$, $\sigma_{\textsc{bp}}^+=0.20\,c/\ell_{\rm c}$, and $\sigma_{\textsc{bp}}\propto l^{-0.2}$; the model for $a=3.$ assumes: $k_0(l) = 3.0 + 2(\l/\ell_{\rm c})^2$, other parameters unchanged.

\end{document}